# Apples to Oranges:

# Causal Effects of Answer Changing in Multiple-Choice Exams

Yongnam Kim

*Department of Educational, School & Counseling Psychology, University of Missouri–Columbia*

Email: ykcpb@missouri.edu

October 14, 2019

### ABSTRACT

Whether examinees' answer changing behavior while taking multiple-choice exams is beneficial or harmful is a long-standing puzzle in the educational and psychological measurement literature. Formalizing the problem using the potential outcomes framework, this article shows that the traditional method of comparing the proportions of "wrong to right" and "right to wrong" answer changing patterns—a method that has recently been criticized by van der Linden, Jeon, and Ferrara (2011)—indeed correctly identify the sign of the average answer changing effect, but only for those examinees who actually changed their initial responses. This subgroup effect is referred to as the average treatment effect on the treated (ATT) and generally differs from the average treatment effect on the untreated (ATU), that is, those who did not change their initial responses. Analyzing two real data sets, including van der Linden et al.'s (2011) controversial data, this article finds that the ATT of answer changing is positive while the ATU of answer changing is negative, therefore, the debate on answer changing effects can be easily resolved. The article also shows that answer changing and answer reviewing are two distinct treatments and knowing answer changing effects is not informative for predicting answer reviewing effects.

KEYWORDS: answer changing, answer reviewing, potential outcomes, causal inference, average treatment effect on the treated, multiple-choice exam

### ACKNOWLEDGEMENTS

I thank Felix Elwert, Peter Steiner, Hyunjoo Lee, Seo Young Lee, Yumi Suk, James Wollack, and students of the advanced seminar in educational measurement and statistics at the University of Wisconsin-Madison for helpful comments. Especially, I thank James Wollack for providing access to his data set for the illustration.

## INTRODUCTION

During multiple-choice exams, examinees frequently find themselves in situations where they are unsure if their first choices are correct or not. Among students and teachers, it has often been advised that examinees should stick to their initial answers because people believe that the *first thought is probably best* (Lynch & Smith, 1972). This belief is still widespread today (see Liu, Bridgeman, Gu, Xu, & Kong, 2015, for recent examples; see also Kruger, Wirtz, & Miller, 2005, for psychological explanations for this belief). However, test theorists and experts in measurement have found quite opposite results. Since Lehman (1928), empirical studies have consistently claimed that examinees have gains by changing their answers (e.g., Liu et al., 2015; Lynch & Smith, 1972; Mathews, 1929; McMorris, DeMers, & Schwarz, 1987; Pagni et al., 2017; Wainscott, 2016). Benjamin, Cavell, and Shallenberger (1984) reviewed 33 studies and concluded that answer changing is indeed *beneficial* and the popular belief is wrong. However, van der Linden, Jeon, and Ferrara (2011) have recently claimed that their advanced item response theory (IRT) models, which are alleged not to be affected by confounding bias due to examinees' ability levels, found a negative answer changing effect, and supported the popular belief. Since then, the effects of answer changing have gained new attention in the educational and psychological literature (see Bridgeman, 2012; Jeon, Boeck, & van der Linden, 2017; Kievit, Frankenhuis, Waldorp, & Borsboom, 2013; Liu et al., 2015).

While the focus of recent literature has been on performing empirical studies with diverse items and subjects (e.g., Liu et al., 2015) and/or analyzing the data with sophisticated techniques (e.g., Jeon et al., 2017; van der Linden et al., 2011), a causal perspective on answer changing behavior has not been formally discussed in the literature. As a result, what kind of causal effects researchers are looking at and what assumptions are required to identify the effects remain opaque to researchers. Although it is not yet popular in the educational and psychological measurement literature, mathematical frameworks that deal with *causation* (instead of association) were developed and have already been used in many theoretical and practical studies for making causal inferences. The most popular one is the potential outcomes framework, developed by Rubin (1974, 1978), also referred to as the Rubin Causal Model (Holland,1986). This causal framework can shed new light on the puzzle of the answer changing effect.

This article investigates the causal effect of answer changing. Although the literature has been reluctant to make this intention clear, the article emphasizes the fact that we aim to know how answer changing behavior (treatment) *causally* affects examinees' final scores (outcome). With clear definitions of treatment and corresponding potential outcomes, different types of causal effects of answer changing are defined: The average treatment effects (ATEs), the average treatment effects on the treated (ATTs), and the average treatment effects on the untreated (ATUs). It is shown that most of the prior studies have estimated ATTs, which are not generally comparable with ATEs or other types of causal effects. Using potential outcomes, the article provides simple nonparametric formulas to compute the causal effects of answer changing without relying on complex cognitive models and modeling assumptions (e.g., IRT models).

The rest of the article is organized as follows. The next section defines causal effects of answer changing using potential outcomes and predicts the missing potential outcomes based on the special nature of the treatment (i.e., answer changing) under a simple true/false item case. The findings are then extended to a general multiple-choice item case. The subsequent section analyzes two real data sets, including van der Linden et al.'s (2011) controversial data—a set in which they found a contradictory result from prior studies but was later acknowledged to be from a flawed analysis due to misalignment error (see Erratum of van der Linden et al., 2011)—and provides empirical evidence of the heterogeneous answer changing effect. The article concludes with a discussion of the difference between different causal effects of answer changing (i.e., ATEs, ATTs, and ATUs) and the difference between answer changing effects and answer reviewing effects.

<div align="center">

CAUSAL EFFECTS OF CHANGING ANSWERS

</div>

## Defining Causal Estimands

The potential outcomes framework helps us mathematically define target causal quantities regardless of analysis procedures. For examinee $i$, let $T_i$ denote the treatment variable, whether he or she changes the initial answer ($T_i = 1$) or not ($T_i = 0$), and $Y_i$ the outcome, whether the final answer is correct ($Y_i = 1$) or incorrect ($Y_i = 0$), for a single item. Given the treatment and outcome, the *potential treatment outcome* $Y_i(1)$ is defined as the examinee's hypothetical final answer correctness if the examinee would have *changed* the initial answer, and the *potential control outcome* $Y_i(0)$ is defined as the hypothetical final answer correctness if the examinee would have *retained* the initial answer. Note that those potential outcomes can be counterfactual, contrary to fact, and conceptually differ from the actually observed (factual) outcome $Y_i$. Then, the unit-level causal effect of changing initial answers can be easily defined as the difference between the two potential outcomes, $\tau_i = Y_i(1) - Y_i(0)$. The quantity $\tau_i$ represents how much unit $i$ would gain if the unit would have changed his or her initial answer, compared to if the unit would have retained it.

Instead of focusing on a particular unit, researchers frequently aim to identify the *average treatment effect* (ATE) across all units in the population. This can be formally defined as

$$ATE = E[\tau_i] = E[Y_i(1)] - E[Y_i(0)], \tag{1}$$

where the expectation is taken over every unit in the population. Sometimes, policymakers are more interested in the average treatment effect on a specific subpopulation, such as those who actually received the treatment (Heckman, Ichimura, & Todd, 1997). This is referred to as the *average treatment effect on the treated* (ATT) and is formally defined as

$$ATT = E[\tau_i \mid T_i = 1] = E[Y_i(1) \mid T_i = 1] - E[Y_i(0) \mid T_i = 1], \tag{2}$$

where the expectation is only taken over treated units. In the context of this article, it quantifies the causal effect of answer changing only for those who actually changed their initial answers.

Although ATEs and ATTs are the two most popular causal estimands in the literature, depending on the research questions, one may also define other types of causal effects. For example, one may define the average treatment effect on those who do *not* receive the treatment. This effect can be referred to as the *average treatment effect on the untreated* (ATU) and is formally defined as

$$ATU = E[\tau_i \mid T_i = 0] = E[Y_i(1) \mid T_i = 0] - E[Y_i(0) \mid T_i = 0], \tag{3}$$

where the expectation is taken over only untreated or control units. In the answer changing context, it quantifies the causal effect of answer changing only for those who stayed with their initial answers.

## Counterfactuals by Response Types in True/False Items

Before proceeding to the identification of the previously defined causal effects, let's first consider examinees' response types. Given an item, all possible response combinations of initial-final answer correctness are categorized as four types: wrong-wrong (*WW*), wrong-right (*WR*),

TABLE 1.

*Response patterns and corresponding observed and potential variables from true/false questions*

| Type | $F$ | $Y$ | $T$ | $Y(1)$ | $Y(0)$ |
|------|-----|-----|-----|--------|--------|
| *WW* | 0 | 0 | 0 | ( a ) | 0 |
| *WR* | 0 | 1 | 1 | 1 | ( b ) |
| *RW* | 1 | 0 | 1 | 0 | ( c ) |
| *RR* | 1 | 1 | 0 | ( d ) | 1 |

*Note.* $F$ = first answer correctness; $Y$ = final answer correctness; $T$ = answer changing status; $Y(1)$ = potential treatment (i.e., changing first answer) outcome; $Y(0)$ = potential control (i.e., retaining first answer) outcome; *WW* = 'wrong' to 'wrong' response type; *WR* = 'wrong' to 'right' response type; *RW* = 'right' to 'wrong' response type; *RR* = 'right' to 'right' response type.

right-wrong (*RW*), and right-right (*RR*), as presented in Table 1. For ease of exposition, here the simplest possible multiple-choice item—that is, a true/false question (or items having only two alternatives) is considered; a general case is consider later. In Table 1, $F_i$ denotes examinee *i*'s first answer correctness, whether the first answer is correct ( $F_i = 1$ ) or incorrect ( $F_i = 0$ ). Thus, for the group of examinees of the type *WW*, the corresponding first and final answers are straightforward: $F_{i \in WW} = 0$ and $Y_{i \in WW} = 0$. Their treatment status is also straightforward: since their first answers are identical to the final answers, they must *not* have changed their initial responses in true/false items: $T_{i \in WW} = 0$.[1] Similarly, one can easily verify the triplet (*F*, *Y*, *T*) for the other three response groups. Essentially, only those who belong to either *WR* or *RW* would have changed their initial answers.

Note that the variables *F*, *Y*, and *T* are all observable. In order to investigate the causal effects of answer changing, one should figure out potential outcomes *Y*(1) and *Y*(0), which are basically latent. Fortunately, one of the potential outcomes can be predicted. If examinee *i* was assigned to the treatment condition, the observed outcome would correspond to the potential treatment outcome $Y_i(1)$, while if the examinee was assigned to the control condition, the observed outcome would correspond to the potential control outcome $Y_i(0)$. In the causal inference literature, this principle is referred to as the *consistency* and is formally expressed (Robins, 1986) as

$$Y_i = T_i \times Y_i(1) + (1 - T_i) \times Y_i(0). \tag{4}$$

---

[1] As with many other studies (e.g., Jeon et al., 2017; Liu et al., 2015; van der Linder at al., 2011), this article only considers examinees' first and final answers and ignore multiple corrections such as "wrong to right to wrong."

Therefore, in Table 1, for the groups *WW* and *RR* who did not change their initial answers, the column of $Y(0)$ can be filled up with their observed outcome $Y$. Similarly, for the other groups *WR* and *RW* who changed their initial answers, the column of $Y(1)$ is filled up with their observed outcome $Y$. Unfortunately, however, one of the potential outcomes cannot be filled up by the consistency principle because it is not realized (e.g., potential treatment outcome are not realized for those who do not receive the treatment). Therefore, in Table 1, the potential outcomes in (a), (b), (c), and (d) seem to remain unknown. In the causal inference literature, this missingness problem is referred to as the *fundamental problem of causal inference* (Holland, 1986).

**Predicting Counterfactuals in True/False Items**

Although the fundamental problem of causal inference is generally hard to solve, it does not occur in the context of answer changing in true/false items. The fact that the treatment variable is whether or not one changes his or her initial answer implies that the corresponding potential control outcome $Y(0)$ should be identical to the initial response $F$. This is because the

TABLE 2.

*Substituting the potential outcomes and computing the causal effects from Table 1*

| Type | $F$ | $Y$ | $T$ | $Y(1)$ | $Y(0)$ | $\tau_i$ | $P(.)$ |
|------|-----|-----|-----|--------|--------|----------|--------|
| *WW* | 0 | 0 | 0 | ( 1 ) | 0 | +1 | $P(WW)$ |
| *WR* | 0 | 1 | 1 | 1 | ( 0 ) | +1 | $P(WR)$ |
| *RW* | 1 | 0 | 1 | 0 | ( 1 ) | −1 | $P(RW)$ |
| *RR* | 1 | 1 | 0 | ( 0 ) | 1 | −1 | $P(RR)$ |

*Note.* $F$ = first answer correctness; $Y$ = final answer correctness; $T$ = answer changing status; $Y(1)$ = potential treatment (i.e., changing first answer) outcome; $Y(0)$ = potential control (i.e., retaining first answer) outcome; $\tau_i$ = causal effect, $Y(1) - Y(0)$; $P(.)$ = group proportion; $WW$ = 'wrong' to 'wrong' response type; $WR$ = 'wrong' to 'right' response type; $RW$ = 'right' to 'wrong' response type; $RR$ = 'right' to 'right' response type.

potential control outcome is, by definition, the hypothetical final answer correctness if one would have *retained* the initial response. Thus, the following holds:

$$Y_i(0) = F_i,$$
(5)

which allows us to predict the counterfactual outcome $Y(0)$ using the observed variable $F$ (i.e., initial answer correctness). Also, the potential treatment outcome $Y(1)$ is the hypothetical final answer correctness if one would have *changed* the initial response. Therefore, it should be opposite to the initial answer in true/false items. Thus, the following holds:

$$Y_i(1) = 1 - F_i.$$
(6)

Using Equations (5) and (6), all the remaining missing potential outcomes in Table 1 can be imputed, as presented in Table 2. For example, for the type *WW*, the missing potential treatment outcome in (a) in Table 1 should be opposite to the initial answer by Equation (6):

$Y_{i \in WW}(1) = 1 - F_{i \in WW}$. As $F_{i \in WW} = 0$, the potential treatment outcome is one: $Y_{i \in WW}(1) = 1$. That is, if one's initial answer was indeed incorrect, changing a response to the other choice in a true/false item will make it right. All the missing potential outcomes in (b), (c), and (d) in Table 2 can be filled out in this way. As a result, the causal effect $\tau_i$ defined by the difference between the two potential outcomes is directly computed by each response type: $\tau_{i \in WW} = \tau_{i \in WR} = 1$ and $\tau_{i \in RW} = \tau_{i \in RR} = -1$.

Let $P(.)$, the last column in Table 2, denote the proportion of examinees' response type in the entire population such that $P(WW) + P(WR) + P(RW) + P(RR) = 1$. Then, the ATE can be simply expressed as the weighted average of each group's causal effects:

$$ATE = (+1) \times P(WW) + (+1) \times P(WR) + (-1) \times P(RW) + (-1) \times P(RR)$$

$$= P(WW) + P(WR) - P(RW) - P(RR). \tag{7}$$

Since all the proportions are known to researchers, the ATE is exactly computed. If one is interested in the ATT, instead of the entire population, cases where $T = 1$ are only considered. This is given by

$$ATT = (+1) \times \frac{P(WR)}{P(WR) + P(RW)} + (-1) \times \frac{P(RW)}{P(WR) + P(RW)}$$

$$= \frac{P(WR) - P(RW)}{P(WR) + P(RW)}, \tag{8}$$

which is again computable. Note that the proportion difference in the numerator is scaled by the proportion of those who changed their initial responses, $P(T = 1) = P(WR) + P(RW)$. Similarly, the ATU is computed by

$$ATU = (+1) \times \frac{P(WW)}{P(WW) + P(RR)} + (-1) \times \frac{P(RR)}{P(WW) + P(RR)}$$

$$= \frac{P(WW) - P(RR)}{P(WW) + P(RR)}. \tag{9}$$

Again, all the quantities are observed and thus the ATU is exactly computed. As such, interestingly, under true/false item cases, identification of causal effects of answer changing does not suffer from the fundamental problem of causal inference.

**General Multiple-Choice Items**

The key properties behind the previous derivation, which does not suffer from the fundamental problem of causal inference, are Equations (5) and (6). Obviously, such properties strongly rely on the fact that there are only two alternatives in true/false or 2-choice items: if one is incorrect, then the other must be correct, and vice versa. In the literature on the answer changing effect, however, researchers have frequently considered $k$-choice items, where $k$ denotes the number of alternatives, and $k > 2$ (e.g., McMorris, DeMers, & Schwarz, 1987; Reile & Briggs, 1952; Skinner, 1983; van der Linden & Jeon, 2012). With this type of general multiple-choice items, if one chose an incorrect alternative at first and then changed to another alternative, the final answer may still be incorrect because there are many distractors and only one correct alternative (Throughout this article, it is assumed that there is a single correct answer in multiple-choice items). Consequently, Equation (6), $Y_i(1) = 1 - F_i$, stating that the potential treatment outcome is always opposite to the initial answer correctness, does not generally hold.

Under this situation, it is necessary to separate the examinees belonging to the type *WW* by whether they indeed changed their initial answers or not. That is, even though the initial and final answers were both incorrect, some of the examinees might have retained their initial responses while others might have changed the initial ones to other incorrect alternatives. Thus, as in Table 3, the type *WW* is divided into two subgroups depending on the values of *T*: ($F = 0$, $Y = 0$, $T = 0$) and ($F = 0$, $Y = 0$, $T = 1$). In contrast, for the other three types, there is only one

TABLE 3.

*Response patterns and corresponding observed and potential variables from k-choice items,*
*where* $k > 2$

| Type | $F$ | $Y$ | $T$ | $Y(1)$ | $Y(0)$ |
|------|-----|-----|-----|--------|--------|
| *WW* | 0 | 0 | 0 | ( $a_1$ ) | 0 |
|      | 0 | 0 | 1 | 0 | ( $a_2$ ) |
| *WR* | 0 | 1 | 1 | 1 | ( b ) |
| *RW* | 1 | 0 | 1 | 0 | ( c ) |
| *RR* | 1 | 1 | 0 | ( d ) | 1 |

*Note.* $F$ = first answer correctness; $Y$ = final answer correctness; $T$ = answer changing status; $Y(1)$ = potential treatment (i.e., changing first answer) outcome; $Y(0)$ = potential control (i.e., retaining first answer) outcome; $\tau_i$ = causal effect, $Y(1) - Y(0)$; $P(.)$ = group proportion; *WW* = 'wrong' to 'wrong' response type; *WR* = 'wrong' to 'right' response type; *RW* = 'right' to 'wrong' response type; *RR* = 'right' to 'right' response type.

treatment status and thus no such dividing is necessary. For example, if the first answer was incorrect but the final answer is correct (i.e., *WR*), the examinees must have changed their initial responses. Relying on the consistency in Equation (3), again, half of the potential outcomes in Table 3 can be filled up. But, the other half of the potential outcomes in ($a_1$), ($a_2$), (b), (c), and (d) remains unknown due to the fundamental problem of causal inference.

Although Equation (6) does not hold, the property $Y_i(0) = F_i$ in Equation (5) is still valid even with general $k$-choice items. Regardless of the number of alternatives (whether $k > 2$ or not), the potential control outcomes are the hypothetical final answer correctness had examinees *retained* their initial responses. Therefore, it is possible to impute the missing potential control outcomes in ($a_2$), (b), and (c) in Table 3 with the corresponding first answers $F$: $a_2 = 0$, b = 0, c = 1. However, the missing potential treatment outcome in ($a_1$) cannot be determined. The value can be either 0 or 1. In contrast, since only one alternative is correct, the missing potential treatment outcome in (d) in Table 3 must be zero. If one's first answer was correct, changing it would make it wrong. The imputed result is presented in Table 4, where the subgroup of ($F = 0$, $Y = 0$,

TABLE 4.

*Substituting the potential outcomes and computing the causal effects from Table 3*

| Type | $F$ | $Y$ | $T$ | $Y(1)$ | $Y(0)$ | $\tau_i$ | $P(.)$ |
|------|-----|-----|-----|--------|--------|----------|--------|
| *WW* | 0 | 0 | 0 | ( 1 ) | 0 | +1 | $P(WW_1)$ |
|      | 0 | 0 | 0 | ( 0 ) | 0 | 0 | $P(WW_2)$ |
|      | 0 | 0 | 1 | 0 | ( 0 ) | 0 | $P(WW_3)$ |
| *WR* | 0 | 1 | 1 | 1 | ( 0 ) | +1 | $P(WR)$ |
| *RW* | 1 | 0 | 1 | 0 | ( 1 ) | −1 | $P(RW)$ |
| *RR* | 1 | 1 | 0 | ( 0 ) | 1 | −1 | $P(RR)$ |

*Note.* $F$ = first answer correctness; $Y$ = final answer correctness; $T$ = answer changing status; $Y(1)$ = potential treatment (i.e., changing first answer) outcome; $Y(0)$ = potential control (i.e., retaining first answer) outcome; $\tau_i$ = causal effect, $Y(1) - Y(0)$; $P(.)$ = group proportion; $WW_1$ = 'wrong' to 'wrong' response type whose treatment status is zero and potential treatment outcome is one; $WW_2$ = 'wrong' to 'wrong' response type whose treatment status is zero and potential treatment outcome is zero; $WW_3$ = 'wrong' to 'wrong' response type whose treatment status is one; $WR$ = 'wrong' to 'right' response type; $RW$ = 'right' to 'wrong' response type; $RR$ = 'right' to 'right' response type.

$T = 0$) is further divided into two, depending on the two possible values of ($a_1$). Consequently, the proportion of the type $WW$ is subdivided into three: $P(WW) = P(WW_1) + P(WW_2) + P(WW_3)$.

Since all potential outcomes are imputed, the causal effects of each row in Table 4 are straightforward. The ATE is then the weighted average of such effects:

$$ATE = P(WW_1) + P(WR) - P(RW) - P(RR). \qquad (10)$$

Note that, compared to the ATE for the true/false items in Equation (7), the first term in the right-hand side in Equation (10) becomes a subgroup of the type $WW$. However, this causal effect is not identifiable because the proportion $P(WW_1)$ is unknown because it is defined with the potential outcomes. Although $P(WW)$ and $P(WW_3)$ are known because they can be defined only by the observed variables ($F$, $Y$, and $T$), the other two proportions, $P(WW_1)$ and $P(WW_2)$, are generally unknown. However, using $0 \le P(WW_1) \le P(WW) - P(WW_3)$, one can derive a bound on the ATE:

$$P(WR) - P(RW) - P(RR) \leq ATE$$
$$\leq P(WW) - P(WW_3) + P(WR) - P(RW) - P(RR). \tag{11}$$

In contrast, the causal effect on the treated units, ATT, is *identifiable* because it is given by

$$ATT = \frac{P(WR) - P(RW)}{P(WW_3) + P(WR) + P(RW)}, \tag{12}$$

which consists of all observable proportions. Finally, the causal effect on the untreated cases, ATU, is given by

$$ATU = \frac{P(WW_1) - P(RR)}{P(WW_1) + P(WW_2) + P(RR)}, \tag{13}$$

but is *not* identified because either $P(WW_1)$ or $P(WW_2)$ is unknown. However, again, using $0 \leq P(WW_1) \leq P(WW) - P(WW_3)$, its bound can be computed by

$$\frac{-P(RR)}{P(WW) - P(WW_3) + P(RR)} \leq ATU \leq \frac{P(WW) - P(WW_3) - P(RR)}{P(WW) - P(WW_3) + P(RR)}. \tag{14}$$

In sum, in general $k$-choice items, ATEs and ATUs cannot be identified. But, one can derive a bound on those effects. In contrast, ATTs can always be identified even with the general multiple-choice items.

The ATT formulas in Equations (8) and (12) reveal one interesting finding. In the vast literature on the answer changing effect, researchers have compared the proportions of "wrong to right" and "right to wrong" patterns and concluded that answer changing is beneficial because the "wrong to right" pattern is more frequent than the "right to wrong" pattern (e.g., Benjamin et al., 1984; Edwards & Marshall, 1977; Lynch & Smith, 1972; Skinner, 1983). Although van der Linden et al. (2011) criticized this traditional method, the proportion difference $P(WR) - P(RW)$ is indeed the numerator of the ATT formulas in Equations (8) and (12). Therefore, the traditional method correctly estimates at least the sign of the answer changing

TABLE 5.

*Counts of 5$^{th}$ graders' initial and final responses to math item 1*

|  |  | Final Response | | | |
|---|---|---|---|---|---|
|  |  | A | B | C | D (key) |
|  | A | 3,039 | 13 | 25 | 295 |
| Initial | B | 8 | 1,426 | 27 | 109 |
| Response | C | 26 | 21 | 4,263 | 336 |
|  | D (key) | 37 | 17 | 57 | 60,086 |

*Note*. The total counts here (69,785) is slightly less than the total sample size (69,806) due to students' nonresponses. The missing rate is negligible (less than .05%). For simplicity, this article ignores the missing data issue in the illustration.

effect. It should be noted, however, that the results by the traditional method should be restricted to examinees who changed their initial answers but should not be generalized to the entire population.

## DATA ANALYSIS

### A Statewide 5th Graders' Math Assessment Data

Using the previous formulas, this section investigates the causal effect of answer changing with two real data sets. The first data set is a statewide 5$^{th}$ graders' math assessment data from the common education data set by Cizek and Wollack (2017), in which students' initial and final responses to each of 53 multiple-choice items are recorded. Every item has four alternatives, and the total number of students is 69,806. In Table 5, students' initial and final responses to item #1 are presented. As the key is 'D,' the following equalities hold:

$$P(WW_3) = P(\text{AB}) + P(\text{AC}) + P(\text{BA}) + P(\text{BC}) + P(\text{CA}) + P(\text{CB});$$
$$P(WW) = P(WW_3) + P(\text{AA}) + P(\text{BB}) + P(\text{CC});$$
$$P(WR) = P(\text{AD}) + P(\text{BD}) + P(\text{CD}); \tag{15}$$
$$P(RW) = P(\text{DA}) + P(\text{DB}) + P(\text{DC});$$
$$P(RR) = P(\text{DD}),$$

where $P(\mathrm{AB})$ denotes the proportion of those who first chose 'A' but changed to 'B,' which is

$13/69,785$ in Table 5 (all other proportions are obtained in the same way). Note that $P(WW_3)$

is the proportion of those who first made a wrong answer but changed to another wrong answer,

which is observed.

Applying Equations (11), (12), and (14) to Table 5, it is found that

$$-.85 \leq ATE_{j=1} \leq -.73;$$
$$ATT_{j=1} = .65; \qquad\qquad (16)$$
$$-.87 \leq ATU_{j=1} \leq -.75,$$

where $j$ denotes item index ( $j = 1, 2, \cdots, J$ ). Note that the causal effects of answer changing on item #1 is *heterogeneous* such that the item-specific ATE and ATU are *negative* while the item-specific ATT is *positive*. This pattern generally holds in other math items. In Figure 1, all the item-specific ATEs and ATTs are presented (the bounds on ATUs are similar to that on ATEs).

While no single ATE bound (vertical lines) has a positive lower bound, all ATTs (filled squares)

except three items (i.e., #34, #35, and #46) are positive.

To summarize the result in Figure 1, one can define and compute the test-level answer

changing effects as the average item-specific effects across all the items. The test-level bound on

ATE can be defined as

$$\sum_j lower(ATE_j)/J \leq ATE^{Test} \leq \sum_j upper(ATE_j)/J, \qquad\qquad (17)$$

where $lower(ATE_j)$ and $upper(ATE_j)$ denote the lower and upper bounds of $j$-specific ATE

bound, respectively. The test-level ATU bound is defined in the same way. And, the test-level

ATT is simply defined as the average of all item-specific ATTs:

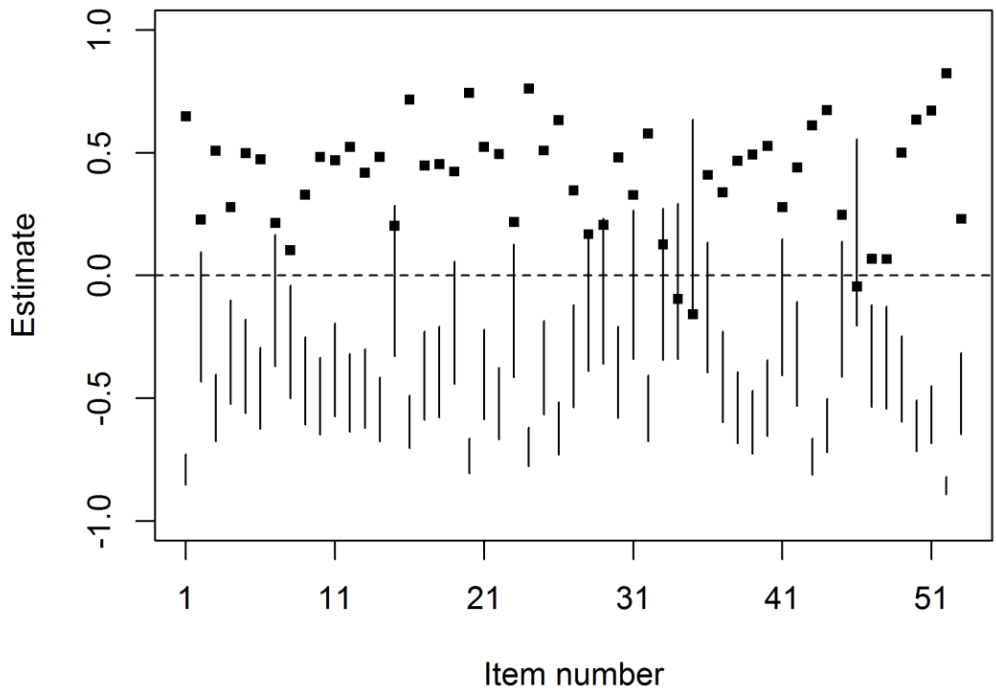

FIGURE 1. *The bounds on the item-specific ATEs, denoted by vertical lines, and the point estimates of the item-specific ATTs, denoted by filled squares, from the statewide math assessment data.*

$$ATT^{Test} = \sum_j ATT_j / J. \qquad (18)$$

It is found that $-.57 \leq ATE^{Test} \leq -.18$, $ATT^{Test} = .40$, and $-.60 \leq ATU^{Test} \leq -.20$. Note that the bound on the test-level ATE is very similar to that on the test-level ATU. This is because the ATE is the weighted average of ATT and ATU, $ATE = P(T=1) \times ATT + P(T=0) \times ATU$, and the answer changing rate, that is, the treatment proportion $P(T=1)$, is rather small in the data set. In the data set, the range of the answer changing rate in each math item was $[1.23\%, 7.76\%]$. Therefore, the ATE becomes closer to the ATU, which has a larger weight.

**Reanalysis of van der Linden, Jeon, and Ferrara's (2011) Data**

The analysis of van der Linden et al.'s (2011) data is at the center of the recent contentious debate on the answer changing effect (see Erratum of van der Linden et al., 2011; Bridgeman, 2012; Jeon et al., 2017; Liu et al., 2015). They criticized the conventional approach implemented by earlier literature, which simply compares the two group proportions, $P(WR)$ and $P(RW)$, and proposed their new IRT approach, alleged not to be affected by confounding bias due to students' ability. Contrary to the *positive* answer changing effect found by the conventional approach, van der Linden et al. (2011) found a *negative* answer changing effect. However, their empirical finding was retracted because they later found a misalignment error in the data files (see their Erratum). From their reanalysis using the corrected data files, van der Linden et al. found a lack of convergence with their proposed IRT approach and thus the conclusion remains inconclusive. Bridgeman (2012) argued that in his own reanalysis using the same data set, he found a *positive* answer changing effect. In the following, this controversial data set is reanalyzed with the causal approach presented in this article.

The data set used by van der Linden et al. (2011) is from a 3<sup>rd</sup> graders' math assessment program, consisting of 65 items and 2,555 students. Unfortunately, item-level data such as the kind presented in Table 5 are not available, but it is confirmed that the contingency table they reported in their article (van der Linden et al., 2011, Table 2, p. 389) is *not* affected by the misalignment error (in personal communication with the authors). In Table 6, their contingency table and the added corresponding proportions are presented. Fortunately, one may still infer the bounds on the ATE, ATT, and ATU from Table 6, where item-specific proportions are collapsed.

The test-level ATE can be expressed as follows:

TABLE 6.

*Counts and proportions (in parentheses) of the four response types in van der Linden et al. (2011, Table 2, p. 389), collected from 65 items and 2,555 examinees*

| | | Final answer | |
| --- | --- | --- | --- |
| | | Wrong | Right |
| | Wrong | 56,587 | 11,543 |
| | | (.34) | (.07) |
| Initial answer | Right | 1,454 | 96,481 |
| | | (.01) | (.58) |

*Note.* The total count here ($56587 + 11543 + 1454 + 96481 = 166065$) is slightly less than $65 \times 2555 = 166075$. This is probably due to nonresponded items, but it is negligible.

$$ATE^{Test} = \sum_j ATE_j / J$$

$$= \left( \sum_j P_j(WW_1) + \sum_j P_j(WR) - \sum_j P_j(RW) - \sum_j P_j(RR) \right) / J \qquad (19)$$

$$= \frac{n(WW_1) + n(WR) - n(RW) - n(RR)}{NJ},$$

where $P_j(.)$ denotes the proportion of the response type with respect to item $j$, $n(.)$ the total number of the response type across all items, $N$ the total number of examinees, and $J$ the total number of items. Although $n(WW_1)$ is not identified, as the latent group count is bounded, $0 \le n(WW_1) \le n(WW)$,[2] and the test-level ATE bound can be bounded:

$$\frac{n(WR) - n(RW) - n(RR)}{NJ} \le ATE^{Test} \le \frac{n(WW) + n(WR) - n(RW) - n(RR)}{NJ}. \qquad (20)$$

Applying Equation (20) to Table 6, it is found $-.52 \le ATE^{Test} \le -.18$, which is negative.

The test-level ATT is expressed as

---

[2] In principle, the upper bound can be further reduced because $n(WW_3)$ can be identified. See Table 4, where the group $WW_3$ can be identified by the observed variables, $F = 0$, $Y = 0$, $T = 1$. Then, one can use $0 \le n(WW_1) \le n(WW) - n(WW_3)$. But, $n(WW_3)$ is not identifiable from Table 6.

$$ATT^{Test} = \sum_j ATT_j \,/\, J$$

$$= \sum_j \left\{ \frac{P_j(WR) - P_j(RW)}{P_j(WW_3) + P_j(WR) + P_j(RW)} \right\} / J \ . \tag{21}$$

$$= \sum_j \left\{ \frac{P_j(WR) - P_j(RW)}{P_j(T=1)} \right\} / J$$

Note that, as opposed to the ATE formulas, the ATT formula above is directly *computable* from the observed data because it only contains the observed groups' proportions (again, note that $WW_3$ is identifiable but $WW_1$ is not; see Table 4). For each item $j$, one just computes the proportion difference between two groups, *WR* and *RW*, and divides it by the proportion of those who changed their initial responses and then, finally, averages across all the $j$-specific ATTs. However, Table 6, which is the contingency table across all 65 items, is not informative enough to compute the test-level ATT using Equation (21). It does not provide any item $j$-specific proportions.

In order to estimate the test-level ATT using Table 6, this article takes a simulation approach. First, as the counts of the subgroups $WW_1$, $WW_2$, and $WW_3$ are unknown, the counts satisfying $n(WW_1) + n(WW_2) + n(WW_3) = n(WW)$ are randomly selected. Second, 2,555 examinees from Table 6 are re-labeled as either $WW_1$, $WW_2$, or $WW_3$ according to the first procedure, and randomly assigned to one of 65 item-specific tables. That is, a pseudo $j$-specific contingency table is randomly created from Table 6 and this process is repeated until all 65 $j$-specific contingency tables are created. Finally, the $j$-specific ATTs from each pseudo $j$-specific table are computed and then their average is computed, which becomes the simulated test-level ATT for a single iteration. The entire procedure is repeated $10^7$ times, and the minimum and maximum values of the simulated test-level ATT bound are derived from the entire repetition. The bound is $.15 \leq ATT^{Test} \leq .78$, which is positive.

The test-level ATU can be obtained by

$$
\begin{aligned}
ATU^{Test} &= \sum_j ATU_j / J \\
&= \sum_j \left\{ \frac{P_j(WW_1) - P_j(RR)}{P_j(WW_1) + P_j(WW_2) + P_j(RR)} \right\} / J \\
&= \sum_j \left\{ \frac{P_j(WW_1) - P_j(RR)}{P_j(T = 0)} \right\} / J
\end{aligned}
\tag{22}
$$

but it is not directly computable since $P_j(WW_1)$ and $P_j(T = 0)$ are unknown. Again, the same simulation approach is applied, implementing the ATU bound formula in Equation (14). The bound is $-1.00 \leq ATU^{Test} \leq -.26$, which is negative. Note that from the same simulation, the test-level ATE bound is also computed, and it was almost identical to the previous analytical result, $-.52 \leq ATE^{Test} \leq -.18$. This supports the validity of the simulation approach. Again, the reason the ATE is closer to the ATU than the ATT is the small proportion of treated units. In van der Linden et al.'s (2011) data, the proportion of the two types, *WR* and *RW*, who obviously changed their responses, is about 7.83%.

In Table 7, the results from both real data sets are summarized. They consistently show the heterogeneous causal effect of answer changing. Changing initial answers is *beneficial* to those who changed their initial answers (ATT) but is *harmful* to those who retained the initial answers (ATU). And, due to the small proportion of examinees who changed their initial answers, the overall causal effect of answer changing (ATE) is closer to ATU.

TABLE 7.

*Test-level ATEs, ATTs, and ATUs from our 5th graders' math assessment data (left) and van der Linden et al.'s (2011) 3rd graders' math assessment data (right)*

|  | 5th graders' math assessment (53 items & 69,806 examinees) | 3rd graders' math assessment (65 items & 2,555 examinees) |
|---|---|---|
| $ATE^{Test}$ | [−.57, −.18] | [−.52, −.18] |
| $ATT^{Test}$ | .40 | [.15, .78] |
| $ATU^{Test}$ | [−.60, −.20] | [−1.00, −.26] |

## DISCUSSION

**Resolving the Answer Changing Debate**

Reile and Briggs (1952, p. 110) asked, "Should students change their initial answers on objective-type tests?" Sixty decades later, Couchman (2015) still asked, "Should you rely on first instincts when answering a multiple choice exam?" While many measurement experts have claimed that changing answers is beneficial (e.g., Benjamin et al., 1984; Liu et al., 2015; Lynch & Smith, 1972; McMorris et al., 1987; Pagni et al., 2017), van der Linden et al. (2011) have claimed that it is harmful. In particular, van der Linden and his colleagues criticized how the vast majority of literature (which has reported positive answer changing effects) does not take into account examinees' ability, which allegedly results in confounding bias in the answer changing effect estimate. However, van der Linden et al.'s proposed sophisticated IRT approach also requires its own modeling assumptions that might not hold in practice (e.g., in their own reanalysis using the corrected data, the model did not converge; see their Erratum). This article provided the causal framework that allows us to formalize answer changing effects irrespective of confounding bias and that does not require any parametric modeling assumptions. The findings reveal that the answer changing debate (see Bridgeman, 2012; Liu et al., 2015; van der Linden et al., 2011) can be resolved by the clear distinction of different causal quantities (e.g.,

ATEs, ATTs, ATUs). Due to the lack of formal causal languages, researchers have previously not realized that each looks at different causal effects, which are not generally comparable.

In the real data analysis, it is found that the answer changing effect is positive, as many prior studies have claimed, to those who changed their initial answers (ATT > 0), but is negative, as van der Linden et al. (2011) claimed, to the entire examinees (ATE < 0) or those who retained their answers (ATU < 0). This indeed makes sense. The examinees who changed their answers did so because they might have a *good reason* to change their original choices (e.g., realizing mistakes). In this case, if answer changing is not allowed (counterfactual), their final score would likely decrease; that is, changing initial answers is beneficial. In contrast, the examinees who retained their initial answers did so because, again, they might have a *good reason* to maintain them (e.g., they were sure about the first solution). In this case, if they were to change their initial responses (counterfactual), the final score would likely decrease; that is, changing initial answers is harmful. The long-standing puzzle may be too obvious if viewed through the causal lens (i.e., the potential outcomes framework). Regardless of whether it involves changing or keeping the first choices, examinees should do whatever they believe to be true.

**The Effect of Answer Reviewing**

Although the question of answer changing itself has long intrigued and puzzled measurement researchers, some researchers consider the question from a more practical perspective. Researchers and test administrators debate whether they should allow examinees to *review* the answered items, especially in computerized adaptive tests (CAT) (see Wise, 1996). Note *whether an examinee has a reviewing opportunity or not* (i.e., answer reviewing) generally differs from *whether an examinee changes or retains his or her initial response* (i.e., answer changing). The relationship between the two behaviors and the final scores can be represented as in a causal diagram in Figure 2. The graph describes that examinees' characteristics (e.g., ability, test anxiety) affect their answer reviewing decision, and upon reviewing, examinees may or may not change their initial answers. Based on their negative answer changing effect, van der Linden et al. (2011) suggested that "it may not be necessary to provide an *opportunity to review* and change answers to previous items in CAT because little may be gained and much risked (emphasis added, p. 396)." This suggestion assumes that knowing the answer changing effect would be informative to predict the answer reviewing effect. Although they made the opposite suggestion (i.e., answer reviewing should be allowed in CAT), Liu et al. (2015) assumed the same rationale.

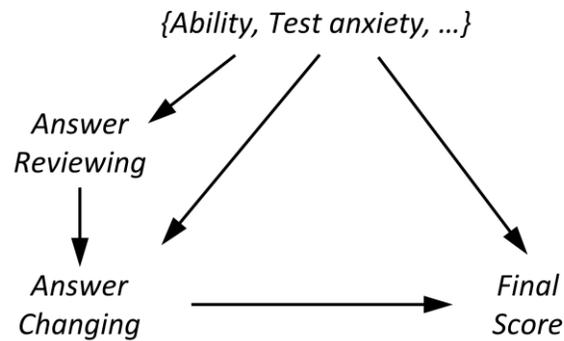

FIGURE 2. *The cognitive process of test taking in multiple-choice exams. Various individual characteristics (e.g., ability, test anxiety, etc.) affect examinees' answer reviewing decision and answer changing behavior. Importantly, the answer changing behavior is followed by the answer reviewing decision.*

Although it sounds reasonable, such rationale is not generally true. The answer *changing* effect that many researchers have investigated, including this article, is *not* useful in knowing the answer *reviewing* effect, which might be of more practical interest. Pearl (2014) provided a useful theorem for this seemingly unnatural relationship. According to his theorem (Theorem 2, p. 110), inferring the causal effect of *A* on *Y* from the other treatment *B*'s effect on *Y* strictly requires that *B* affects *Y* only via *A* (i.e., all directed paths from *B* to *Y* go through *A*). Otherwise, the causal effect of *A* on *Y* cannot be inferred from the causal effect of *B* on *Y*. Replacing *A* with *Answer Reviewing* and *B* with *Answer Changing*, one can see that the causal effect of answer reviewing cannot be inferred from the causal effect of answer changing because as Figure 2 clearly shows, *Answer Reviewing* does not sit on the direct path of *Answer Changing* → *Final Score*. When it comes to the *signs*, rather than the magnitudes, of causal effects, the same is true. Knowing the sign of the average effect of answer reviewing on answer changing (*Answer Reviewing* → *Answer Changing*) and the sign of the average effect of answer changing on the final score (*Answer Changing* → *Final Score*) does not allow us to infer the sign of the average effect of answer reviewing on the final score (*Answer Reviewing* → *Final Score*). VanderWeele and Robins (2010) showed that in a causal chain relationship of $A \rightarrow B \rightarrow Y$, when both the signs of the average effect of *A* on *B* and of the average effect of *B* on *Y* are positive, the average effect of *A* on *Y* can even be negative (see their Example 2 in pp. 119-120). In general, one cannot infer the sign of the effect of *A* on *Y* from the signs of *A* on *B* and of *B* on *Y*. Such inference requires a strong assumption such as monotonic effects (see VanderWeele & Robins, 2010, for more details).

Therefore, our knowledge about the answer changing effect does not help us to infer the answer reviewing effect. Although this article found that average answer changing effect for the entire population is negative (ATE < 0), it is possible that the average answer reviewing effect for the same population is positive. Having solved the long-standing puzzle about the answer changing effect, this article does not provide any evidence for the answer reviewing effect. Neither do other studies about answer changing. To answer whether answer reviewing is beneficial or harmful, a separate randomized experiment, manipulating examinees' reviewing status, will be necessary (e.g., Vispoel, 2000).